\documentclass[referee]{aa} % for a referee version
\usepackage{graphicx}
\usepackage{txfonts}
\begin{document}
   \title{Detection of CO in Triton's atmosphere and the nature of surface-atmosphere interactions}
%
%   \subtitle{I. Overviewing the $\kappa$-mechanism}

   \author{E. Lellouch
          \inst{1}
	  \and
          C. de Bergh\inst{1}
          \and
          B. Sicardy\inst{1,2}
	  \and
          S. Ferron\inst{3}
	  \and
	  H.-U. K\"aufl\inst{4}
         }

   \institute{LESIA, Observatoire de Paris, 5 place Jules Janssen, 92195 Meudon, France\\
              \email{emmanuel.lellouch@obspm.fr}
\and 
Universit\'e Pierre et Marie Curie, 4 place Jussieu, F-75005 Paris, France;  senior member of the Institut Universitaire de
France
         \and	
	ACRI-ST, 260, Route du Pin Montard, BP 234, 06904 Sophia-Antipolis Cedex, France
	 \and
             European Space Observatory, Karl-Schwarzschild-Strasse 2, D-85748 Garching bei M\"unchen, Germany  
             }

   \date{Received March, 1, 2010; revised March, 10, 2010; accepted, XX}

% \abstract{}{}{}{}{} 
% 5 {} token are mandatory
 
  \abstract
{}
   {Triton possesses a thin atmosphere, primarily composed of nitrogen, sustained by the sublimation of surface ices}
   {The goal is to determine the composition of Triton's atmosphere and to constrain the nature of surface-atmosphere interactions.}
   {We perform high-resolution spectroscopic observations in the 2.32-2.37 $\mu$m range, using CRIRES at the VLT.}
   {From this first spectroscopic detection of Triton's atmosphere in the infrared, we report (i) the first observation of gaseous methane since its discovery in the ultraviolet by Voyager in 1989 and (ii) the first ever detection of gaseous
CO in the satellite. The CO atmospheric abundance is remarkably similar to its surface abundance, and appears to be controlled by a thin, CO-enriched, surface veneer resulting from seasonal transport and/or atmospheric escape. The CH$_4$ partial pressure is several times larger than inferred from Voyager. This confirms that Triton's atmosphere is seasonally variable and is best interpreted by the warming of CH$_4$-rich icy grains as Triton passed southern summer solstice in 2000. The presence of CO in Triton's atmosphere also affects its temperature, photochemistry and ionospheric composition. An improved upper limit
on CO in Pluto's atmosphere is also reported.}
   {}

   \keywords{Solar system:general ; Infrared: solar system ; Triton}  
 
\titlerunning{Detection of CO in Triton's atmosphere} 
   \maketitle
%
%________________________________________________________________

\section{Introduction}
Like  Pluto, Neptune's satellite and probably former Kuiper-Belt object Triton possesses a tenuous, predominantly nitrogen atmosphere, 
in equilibrium with surface ices mostly composed of N$_2$ and a variety of other species. The most volatile of these species, 
CH$_4$ and CO, must be present in trace amounts in the atmosphere as well. However, depending on the precise mechanisms of surface-atmosphere
interactions, the expected atmospheric abundances vary by orders of magnitude, and except for the detection of CH$_4$ in the UV by 
Voyager in 1989, observations have been severely lacking. Progress in IR-detector technology makes the remote study of thin and distant
atmospheres now possible. Following our observations of methane in Pluto's atmosphere (Lellouch et al. 2009), we here report on the first 
spectroscopic detection of Triton's atmosphere in the infrared.

\section{VLT/CRIRES observations and CH$_4$ and CO measurements }
Spectroscopic observations of Triton were obtained on July 4, 2009, using the CRIRES infrared echelle spectrograph (K\"aufl
et al. 2004) installed on ESO VLT (European Southern Observatory Very Large Telescope) UT1 (Antu) 8.2 m telescope. We focussed on the regions of the (2-0) band of carbon monoxide and of the $\nu_3$+$\nu_4$ band of methane, covering the 2318-2330, 2334-2345, 2349-2359 and 2363-2373 nm ranges. We used the instrument in adaptive optics mode and with a slit of 0.4", providing a mean spectral resolution of 60,000, and acquired spectra during $\sim$ 4 hours. A large Doppler shift (-23 km/s) ensured proper separation of the target lines from the telluric absorptions.

%__________________________________________________________________

%\section{Inferences on Pluto's lower atmosphere structure and methane abundance}

%                                     Two column figure (place early!)
%______________________________________________ Gamma_1 (lg rho, lg e)
The resulting spectrum shows the detection of many lines due to methane in Triton's atmosphere, particularly at 2320-2330 nm (Fig. 1). This is the first observation of gaseous methane since its discovery by Voyager (Herbert and Sandel 1991). As for our study of Pluto's CH$_4$, we constructed a direct line-by-line atmospheric model of Triton, integrated over angles and including solar lines  
reflected off Triton's surface as well as the telluric transmission (see details in Lellouch et al. 2009).
%(calculated by using LBLRTM and checked against observations of a telluric standard) 
 %accounting for their proper relative Doppler shifts. 
The spectrum was first modelled by assuming a single-temperature layer, with Triton's atmospheric methane mean temperature (T) and column density (a) as free parameters.  We inferred T=50$^{+20}_{-15}$~K and a = 0.08$\pm$0.03 cm-am (Fig. 2 on-line).  The same analysis for Pluto  had given T=90$^{+25}_{-18}$~K and a = 0.75$^{+0.55}_{-0.30}$ cm-am. This confirms that Pluto's atmosphere is warmer than Triton's, as a result of its higher methane abundance. 

   \begin{figure}
   \centering
   \includegraphics[width=10cm,angle=270]{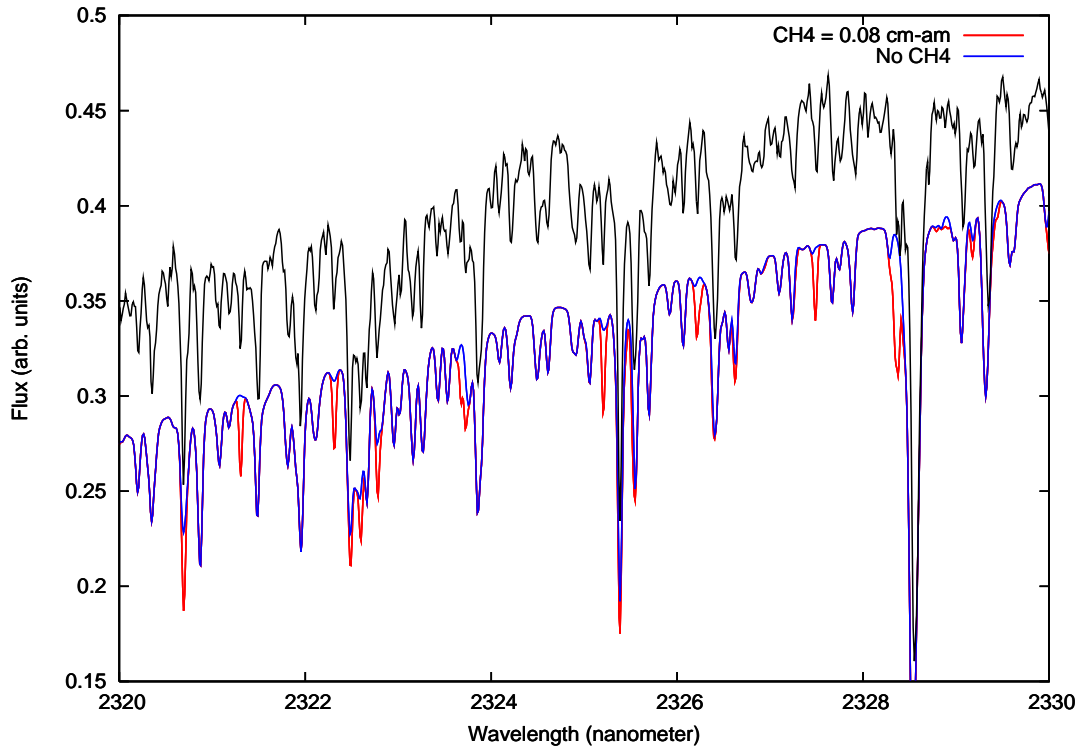}
   \caption{Black: Triton spectrum at 2320 - 2330 nm observed by CRIRES/VLT. The spectral resolution is 60,000. Red and blue curves show synthetic spectra, including telluric and solar lines. Red: methane column density in Triton's atmosphere = 0.08 cm-am. Triton's thermal profile, based on Voyager measurements, is taken from Krasnopolsky (1993) and a Voyager-like vertical distribution is used for methane (Herbert and Sandel 1991, entrance profile). Blue: no methane. The continuum slope is due to the red wing of the $\nu_3$ +  $\nu_4$ band due to solid methane. Here, as in Fig. 3, the vertical unit approximately represents the geometric albedo (but uncorrected for telluric and solar lines). Models are shifted vertically by -0.07 for clarity.}
              \label{FigCH$_4$}%
    \end{figure}

The error bars on the inferred mean methane temperature are such that it is not possible to constrain the methane vertical distribution. Instead, we used the Voyager-determined thermal structure (temperature vs altitude, Krasnopolsky et al., 1993) and methane vertical profile (Herbert and Sandel 1991, ingress UV occultation profile). The latter shows a decrease of the CH$_4$ mixing ratio with altitude with a scale height of $\sim$20 km, due to photolysis. We obtained the same column density as above, indicating a partial pressure of methane of 9.8$\pm$3.7 nbar, i.e. a surface density of (1.9$\pm$0.7) x 10$^{12}$ cm$^{-3}$. This appears to be 4$^{+5}_{-2.5}$ times larger than inferred from Voyager in 1989, adopting the CH$_4$ number densities of Herbert and Sandel (1991) and Strobel and Summers (1995) (4.7$\times$10$^{11}$ cm$^{-2}$, within a factor 1.7, averaging ingress and egress). An even larger enhancement factor (5$^{+6}_{-2}$) is indicated if the Krasnopolsky and Cruikshank (1995) reanalysis of the Voyager UV data, giving CH$_4$ = 3.1$\pm$0.8$\times$ 10$^{11}$ cm$^{-3}$ at the surface, is used. Results are independent on the surface pressure, as collisional broadening is negligible. They clearly demonstrate that the CH$_4$ partial pressure has increased in the last 20 years.

The 2335-2365 nm part of the Triton spectrum (see excerpts in Fig. 3) shows the detection of 8 lines due to the CO(2-0) band (R2-R5, P2, P3, P5 and P8), providing the first detection of CO in its atmosphere. An accurate determination of the CO abundance is particularly difficult, as at infinite spectral resolution, these features are very narrow, saturated Doppler-shaped lines. Nonetheless, assuming a vertically uniform CO distribution, and utilizing the whole set of CO lines (see Fig. 4
on-line), we determine a CO column of 0.30 cm-am, i.e. a CO partial pressure of 24 nbar, within a factor of 3. The column density CO/CH$_4$ ratio is nominally $\sim$3.75 (surface partial pressure ratio CO/CH$_4$ $\sim$2.5), with a factor of 4 uncertainty. Deriving the CO/N$_2$ and CH$_4$/N$_2$ mixing ratio is complicated by the fact that the surface pressure in 2009 is unknown. Stellar occultation results (Olkin et al. 1997, Sicardy et al. 1998, Elliot et al. 1998, 2000a) indicate that the pressure has been doubling in $\sim$ 10 years from the 14 $\mu$bar value determined by Voyager in 1989 (Gurrola 1995). A reasonable assumption for 2009 is 40 $\mu$bar, providing CO/N$_2$ $\sim$ 6$\times$10$^{-4}$ and CH$_4$/N$_2$ $\sim$2.4$\times$10$^{-4}$ at the surface, within factors of 3 and 1.4 respectively. The CO abundance we determine is many times less than previous upper limits (Broadfoot et al. 1989, Young et al. 2001).

\onlfig{2}
{
  \begin{figure*}
   \centering
   \includegraphics[angle=270,width=16cm]{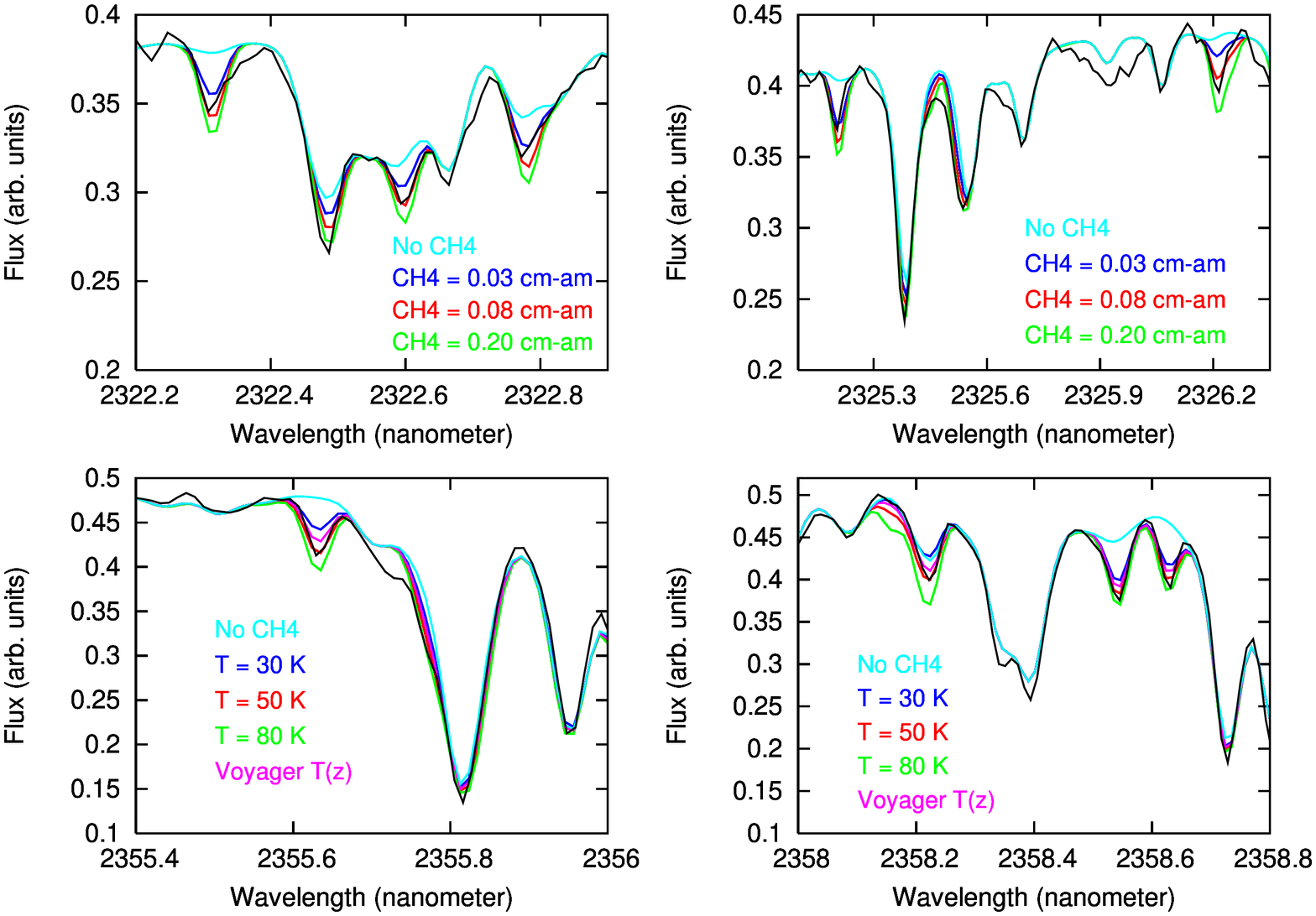}
      \caption{ Zoom on several CH$_4$ lines showing sensitivity of the spectrum to Triton's methane. Some portions of the spectrum, especially over 2320-2326 nm, are relatively independent on temperature, while high-energy lines at 2353-2359 nm show increased temperature sensitivity.The top two panels show sensitivity to the methane abundance. Blue, red, and green synthetic spectra have 0.03, 0.08, and 0.20 cm-am of methane. Triton's thermal profile is taken from Krasnopolsky (1993) and a Voyager-like vertical distribution is used for methane (Herbert and Sandel 1991, entrance profile). Based on these models, the best fit methane column density is determined to be 0.08$\pm$0.03 cm-am (i.e. $\pm$40 \%). The bottom two panels show sensitivity to methane temperature. The previous best-fit model using Voyager thermal profile is shown in pink. Other models assume an isothermal atmosphere with temperature of 30 K (dark blue), 50 K (red) and 80 K (green). These fits indicate a mean methane temperature of 50$^{+20}_{-15}$~K.}
{\it Figure available on-line.}  

         \label{FigallCH$_4$}
   \end{figure*}
}

%__________________________________________________ One column table
%

   \begin{figure}
   \centering
   \includegraphics[width=10cm,angle=270]{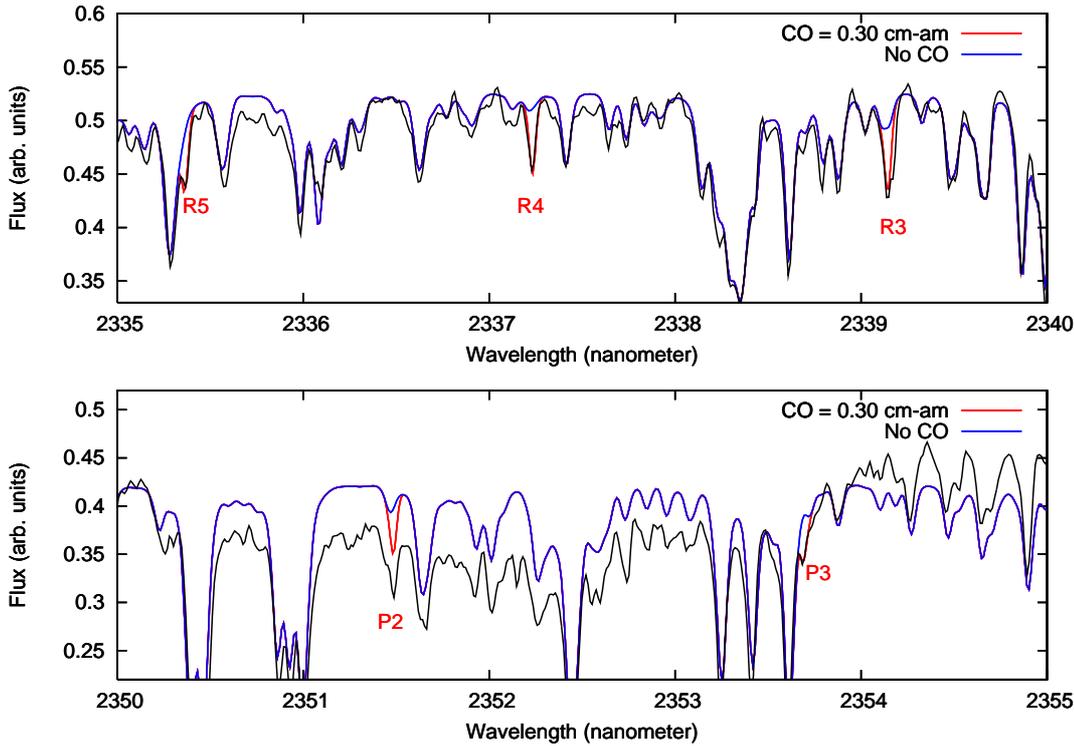}
      \caption{Black: portions of the Triton spectrum at 2335 - 2340 nm and 2350-2355 nm. Five CO lines (R5, R4, R3, P2 and P3) are present in these spectral ranges. Red and blue curves show synthetic spectra, including telluric and solar lines, as well as Triton's methane. Red: CO column density in Triton's atmosphere = 0.30 cm-am. Blue: No CO. The mismatch in the ``continuum" level over 2351-2354 nm is due to the absorption of the (2-0) band of CO ice (see e.g. Quirico et al. 1999, Grundy et al. 2010), not included in models.}
         \label{FigCO}
   \end{figure}

\onlfig{4}
{
  \begin{figure*}
   \centering
   \includegraphics[angle=270,width=16cm]{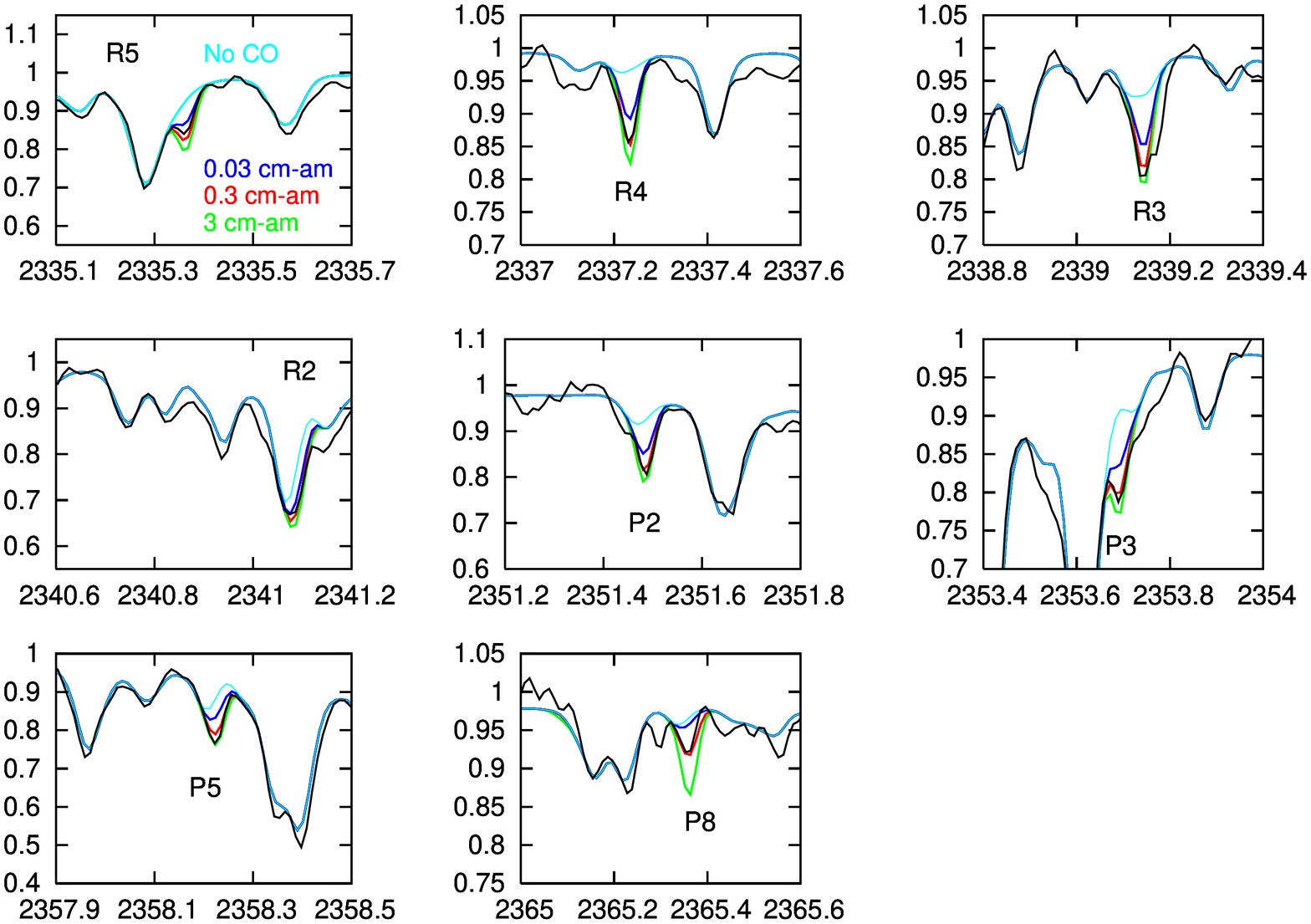}
      \caption{The eight detected CO lines from Triton's atmosphere. X-axis units are nm and Y-axis units are arbitrary. Lines are compared with models including 0, 0.03, 0.30 and 3 cm-am of CO. The slow change of absorption depth with change of column density is caused by heavy saturation of particularly narrow Doppler-shaped lines at T $\sim$50 K. Based on these models, the best fit CO column density is determined to be 0.30 cm-am with a factor of 3 uncertainty.}
{\it Figure available on-line.}  

         \label{FigallCO}
   \end{figure*}
}

%                                                One column figure
%----------------------------------------------------------- S_vib

\section{Discussion}
Near-infrared observations indicate that CO and CH$_4$ are present on Triton's surface with mixing ratios of 0.05 \% and 0.1 \% relative to N$_2$, and at least for CH$_4$, mostly in solid solution in N$_2$ (Cruikshank et al. 1993, Quirico et al. 1999, Grundy et al. 2010). In this situation, the expected partial pressure of each species is the product of its solid mole fraction and its pure vapor pressure (Raoult's law for an ideal mixture). This scenario leads to atmospheric abundances of CO and CH$_4$ that are about 1 and 3 orders of magnitude lower than observed, respectively (Fig. 5). Although Henry's law may be more applicable than Raoult's in the case of the N$_2$-CH$_4$ system, this does not reduce the discrepancy by more than a factor of 2-3. This problem has been studied in the case of Pluto's atmospheric methane, present at the 0.5 \% level (Young et al. 1997, Lellouch et al. 2009), i.e. also considerably enriched over its solid solution equilibrium value. The origin of such enhancement is thought to ultimately lie in the seasonal evolution of the N$_2$-dominated solid solution. Preferential sublimation of N$_2$ initially creates a thin surface layer enriched in the less volatile species. Further evolution of this layer may lead either (i) to the formation of chemically pure grains in vertically or geographically segregated deposits (``pure ice" scenario), or (ii) to the establishment of a homogeneous ``detailed balancing layer" controlling the surface-atmosphere exchanges. In the ``pure ice" case, the atmospheric mixing ratios are in simple proportion of the pure vapor pressures at the relevant ice temperatures (which may be different for different species) and, except for the main species which controls the pressure, of the fractional area covered by each ice. Focussing on the case of CH$_4$ on Pluto, Stansberry et al. (1996) demonstrated that pure CH$_4$ lag deposits (whose existence is proved by observations, see Dout\'e et al. 1999) assume higher temperatures than N$_2$ due to their reduced sublimation cooling and preferential formation in regions of high insolation. Even if covering only a few percent of Pluto's surface, such patches can explain the observed atmospheric abundance of methane.  
Alternatively, the ``detailed balanced" model (Trafton, 1990; Trafton et al. 1998) predicts that surface-atmosphere exchanges in presence of atmospheric escape and seasonal transport lead to an atmospheric composition reflecting that of the accessible ice reservoir from which it is replenished. When no fractionation (e.g. diffusive) occurs during escape or transport, the atmospheric mixing ratios are identical to those in the volatile reservoir. This is accomplished by the thin surface veneer enriched in the less volatile species, throttling off the N$_2$ sublimation, and in permanent equilibrium with the atmosphere according to Raoult's law. 

   \begin{figure}
   \centering
   \includegraphics[width=10cm,angle=270]{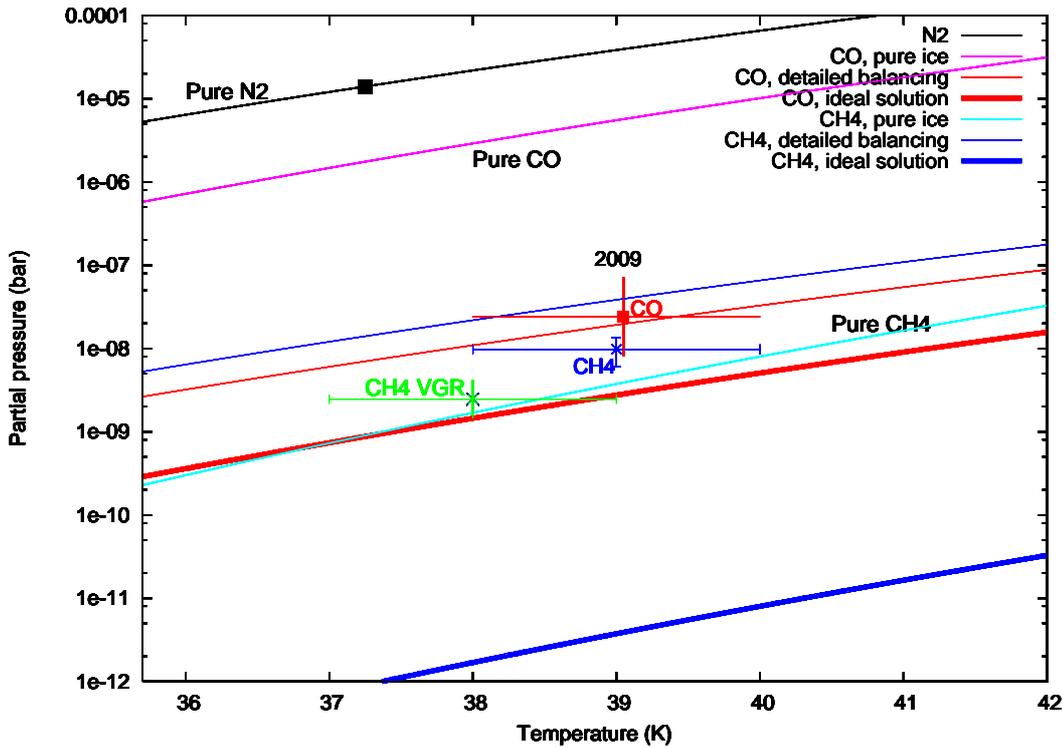}
      \caption{ Abundance measurements and equilibrium curves for Triton's volatiles. The black, pink and light blue curves show the vapor pressure of pure N$_2$, CO and CH$_4$ ices, calculated from Fray and Schmitt (2010). For CO and CH$_4$, the thick curves (red and blue, respectively) show the partial pressures based on Raoult's law for an ideal solid solution with N$_2$, with CO and CH$_4$ respective abundances of 5~x~10$^{-4}$ and 1~x~10$^{-3}$ in the ice (Quirico et al. 1999). For CO and CH$_4$, the thin red and blue show the partial pressures expected in the framework of the ``detailed balancing model" (see text). The black square shows the 1989 Voyager pressure measurement (14 nbar), which corresponds to sublimation equilibrium of N$_2$ ice at 37.25 K. The green symbol represents the Voyager-measured CH$_4$ partial pressure (Herbert and Sandel 1991); it is plotted at
38$\pm$1 K, the surface temperature measured by Voyager and the N$_2$ ice temperature inferred by Tryka et al. (1994). The blue star and the red square show the CH$_4$ and CO abundances determined in this work. They are plotted at a temperature of 39$\pm$1 K. A 39 K temperature corresponds to a plausible 40 $\mu$bar pressure for Triton's atmosphere in 2009; 40 K, which corresponds to a N$_2$ vapor pressure of 65 $\mu$bar, is a reasonable upper limit, given the stellar occultation inferences that Triton's surface pressure has doubled in the 10 years following the Voyager epoch. The CO partial pressure
we measure is consistent with expectations from the detailed balancing model, while CH$_4$ is lower.
              }
         \label{Figeq}
   \end{figure}

   \begin{figure}
   \centering
   \includegraphics[width=10cm,angle=0]{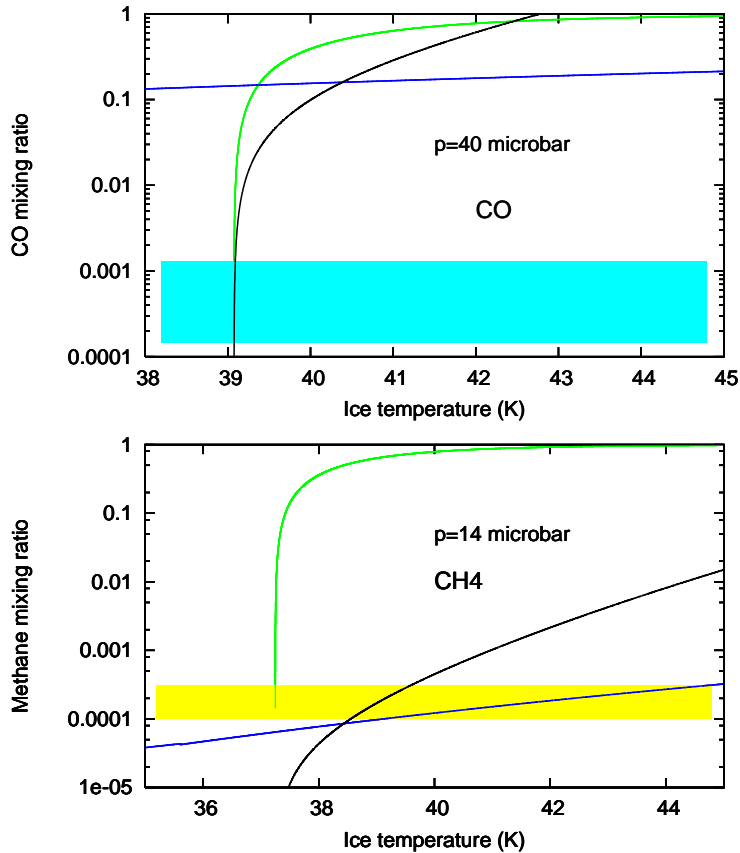}
      \caption{ Atmospheric mixing ratios and composition of the ice boundary layer (``film") in the detailed balancing model. (Top): CO-N$_2$ system. A surface pressure of 40 $\mu$bar is assumed (estimate for 2009). The dark blue line is the CO/N$_2$ mixing ratio expected for pure ices. The green curve shows the CO mole fraction in the ice surface film, and the black curve is the CO/N$_2$ atmospheric mixing ratio derived from the composition of the ice film and applying Raoult's law. The range of CO/N$_2$ atmospheric mixing ratios inferred in this work for this pressure ((2-18) x 10$^{-4}$) is indicated by the  blue-colored region. It implies a CO/N$_2$ mixing ratio in the surface film of (1.4-12) x 10$^{-3}$ (see text). The surface film is therefore largely dominated by N$_2$, and the total pressure is defined by N$_2$ equilibrium at 39.075 K.
(Bottom): CH$_4$-N$_2$ system.  Calculations are here performed for a 14 $\mu$bar pressure, appropriate for the Voyager conditions. The Voyager-determined CH$_4$/N$_2$ atmospheric mixing ratio at the surface level ((1.1-3) x 10$^{-4}$) is indicated by the yellow region. The colored lines have the same meaning than in the top panel, with CH$_4$ replacing CO. The intersection of the black line with the colored area shows that explaining the observed CH$_4$ mixing ratio and the total pressure requires a 38.3-39.6 K surface temperature and would imply a very high CH$_4$ mole fraction (50-80 \%) in the surface film, well beyond the solubility limit of CH$_4$ in N$_2$. The same conclusion is reached if the CH$_4$ amounts measured in 2009 are used. Note that these diagrams remain similar at other surface pressures, the only change being the required ice temperature to sustain the total pressure.
              }
         \label{Figdb}
   \end{figure}

Our observations provide discriminating keys on these scenarios. The case for CO is most straightforward because (i) CO is not subject to diffusive separation from N$_2$ upon escape (ii) as the ratio of its vapor pressure to that of N$_2$ is largely insensitive to temperature in the relevant range (e.g. CO/N$_2$ = 0.112 at 36 K and 0.166 at 41 K), its atmospheric mixing ratio should be roughly seasonally constant. The observed CO abundance is over two orders of magnitude lower than the pure CO vapor pressure (Fig. 5), and one might envisage that the atmospheric CO results from the sublimation of pure CO patches covering $\sim$0.4~\% of the surface. However, we do not regard this scenario as likely. Although the pure vs isolated form of CO on Triton's surface has not yet been proven from observations, the miscibility of CO and N$_2$ in solid phase in all proportions and the similarity of their vapor pressures argue for a co-condensation of the two species on Triton's surface. This is further supported by the strikingly similar longitudinal distribution of the N$_2$ and CO ice bands at Triton (Grundy et al. 2010), strongly suggestive of a spatially constant CO/N$_2$ ice mixing ratio. In addition, even if pure CO patches occurred on Triton's surface, they would probably not be able to elevate the CO atmospheric abundance along the mechanism envisaged for Pluto methane. This is because CO is not buoyant in N$_2$, restricting the dispersal of CO-rich gas in the background atmosphere and therefore the sublimation rates of the CO patches (Stansberry et al. 1996). Instead, the detailed balancing model provides a physically-expected interpretation to the fact that the atmospheric CO/N$_2$ mixing ratio is consistent with its value in the ice phase (Fig. 5). Based on this scenario, the N$_2$-CO composition of the surface boundary layer (``film") can be established by simple application of Raoult's law, providing q$_{CO}$(film) = q$_{CO}$(atm)$\times$p$_{sat}$(CO) / p$_{sat}$(N$_2$) (Trafton et al. 1998). Adopting again p = 40 $\mu$bar, our observed q$_{CO}$(atm)  =  (2--18)$\times$ 10$^{-4}$ indicates q$_{CO}$(film)= (1.4--12)$\times$10$^{-3}$. Therefore the surface veneer is still dominated by N$_2$ and the presence of CO does not importantly modify the N$_2$ atmospheric pressure, defined by equilibrium at 39.1 K (Fig. 6). Because it may be as thin as a few molecular layers, the surface film may not be visible in the near-IR spectra.

The case for CH$_4$ is more complex. As previously realized (Cruikshank et al. 1993, Yelle et al. 1995, Strobel and Summers 1995, Strobel et al. 1996), the CH$_4$ atmospheric mixing ratio at the surface measured by Voyager ($\sim$1.8$\times$10$^{-4}$) is at least three orders of magnitude larger than expected for ideal mixture. However, we note that it is also smaller, by a factor of $\sim$6, than the ice CH$_4$/N$_2$ mixing ratio, and as such does not agree with the detailed balancing model in its simplest form. Unlike CO, CH$_4$ is subject to atmospheric photolysis and mass separation, and its vapor pressure is more temperature-dependent. This probably makes the surface/atmosphere abundance relationship for CH$_4$ complex and seasonally variable. In any case, the phase diagram of N$_2$-CH$_4$ is not obviously consistent with the formation of a CH$_4$-rich solid solution veneer (Stansberry et al., 1996). In fact, explaining the range of observed CH$_4$ atmospheric abundance would require a CH$_4$ mole fraction in the surface film as high as 50-80 \% (Fig. 6), well beyond the solubility limit of CH$_4$ in N$_2$ (Prokhvatilov and Yantsevich 1983).%,Lunine and Stevenson 1985). 
The formation of pure CH$_4$ ice grains, decoupled from the mixture and not influencing its sublimation (Stansberry et al. 1996, Spencer et al. 1997), further evolving into a lag deposit, may be a more plausible outcome. Using a Bond albedo of 0.85 (Triton's polar cap) and an emissivity of 0.7-1, a reasonable subsolar temperature for these pure methane patches is 45-48 K. Applying the Stansberry et al. (1996) Pluto model then indicates that methane patches covering 0.5-1 \% of Triton's surface are sufficient to maintain a $\sim$2$\times$10$^{-4}$ atmospheric mixing ratio. Although there is no evidence for such patches in Triton's near-IR spectrum, the methane longitudinal distribution of CH$_4$ ice is different from that of N$_2$, and small areas of CH$_4$-dominated ice, notably near 300$^{\circ}$ longitude, are not inconsistent with observations (Grundy et al. 2010). In contrast, the existence of widespead pure methane ice is ruled out; therefore the emphasized fact that the Voyager-measured methane partial pressure was consistent with vapor pressure equilibrium of pure CH$_4$ ice at 38 K is probably coincidental.

After the Voyager encounter, a variety of seasonal N$_2$ cycle models (see review in Yelle et al. 1995) were explored to attempt explaining Triton's visual appearance and the then measured surface pressure. These models, which essentially differed in the assumed ice and substrate albedos and thermal inertia, had limited success, leaving unanswered the simple question of where the ice is on Triton. Yet, they made distinctive predictions as to the short-term evolution of Triton's atmosphere. High thermal inertia models predicted a pressure increase as Triton approached and passed Southern summer solstice in 2000 (Triton subsolar latitude moved from 45.5 S in 1989 to a maximum 50 S in 2000 and 47 S in 2009). This is a consequence of increased insolation on, and attendant sublimation of, the Southern polar cap (Spencer and Moore 1992, Forget et al. 2000). In contrast, ``dark frost" models (Hansen and Paige 1992) or low thermal inertia models predicted a pressure decrease from $\sim$1980 on, due to the exhaustion of the seasonal southern cap and re-condensation of N$_2$ on the invisible winter pole. The discovery of the pressure increase in the 1990's, and the persisting signature of N$_2$ and other ices in Triton reflectance spectrum with no obvious temporal evolution (Grundy et al. 2010), strongly argue for the fact that the bright deposits covering most of Triton southern hemisphere are indeed relatively stable seasonal deposits. Our observation that the methane partial pressure has increased by a factor $\sim$4 from 1989 to 2009 is qualitatively consistent with the reported pressure increase and the above interpretation. Since the CH$_4$ vapor pressure varies more sharply with temperature than N$_2$, we expect that atmospheric methane is currently increasing more rapidly than pressure, but multi-volatile seasonal models will be needed to fully interpret our results. A direct measurement of Triton's current pressure is also highly desirable, and could be obtained through a redetermination of the N$_2$ ice temperature from its 2.15 $\mu$m band (Tryka et al. 1994).

The detection of CO has also implications on Triton's atmospheric thermal structure, photochemistry, and ionosphere. CO is an important cooling agent through radiation in its rotational lines (Krasnopolsky et al. 1993, Strobel and Summmers 1995, Elliot et al. 2000b). It enriches atmospheric chemistry by introducing additional oxygen species (Krasnopolsky and Cruikshank 1995). Most importantly, it profoundly modifies ionospheric composition by providing a source of C atoms and C$^+$ ions %(Delistsky et al. 1992, Lyons et al. 1993) 
and by suppressing the N$^+$ concentration %(Lellouch et al. 1992) 
at the benefit of CO$^+$ and NO$^+$ (see review in Strobel and Summers 1995). 
%All these aspects led Krasnopolsky and Cruikshank (1995) to regard direct measurements of atmospheric CO as ``badly needed". 
Although the error bar on the CO abundance is still large, all previous considerations on the role of CO have now direct observational support.  

During the same observing night, we also searched for CO in Pluto's atmosphere, covering the region of the (3-0) band near 1.57 $\mu$m. Only an upper limit (1 cm-am) was obtained. For a characteristic surface pressure of 15 $\mu$bar (Lellouch et al. 2009), this indicates CO/N$_2$ $<$ 5$\times$10$^{-3}$. While improved over previous results (Bockel\'ee-Morvan et al. 2001, Young et al. 2001), this upper limit is still relatively unconstraining when compared to the measured CO ice mole fraction (1$\times$10$^{-3}$, Dout\'e et al. 1999). Nonetheless, given the similarity of the two bodies, the above considerations on the surface control of atmospheric CO at Triton should apply to Pluto as well. In 2015, observations with the ALICE and Rex instruments on New Horizons will provide measurements of the surface pressure and CH$_4$ and CO abundance in Pluto's atmosphere. We anticipate that CO will be measured at a uniform ratio of 0.001 and that the methane mixing ratio will show horizontal variability associated with local time and methane patch distribution.

\begin{acknowledgements}
This work is based on observations performed at the European Southern Observatory (ESO), proposal 383.C-0609. We thank F. Forget, B. Schmitt,
and L. Trafton for illuminating discussions.
\end{acknowledgements}


\begin{thebibliography}{}

%  \bibitem[1966]{baker} Baker, N. 1966,
%      in Stellar Evolution,
%      ed.\ R. F. Stein,\& A. G. W. Cameron
%      (Plenum, New York) 333

\bibitem[1989]{broadfoot}Broadfoot, L. et al. 1989, Science,246, 1459

\bibitem[2001]{bock01}Bockel\'ee-Morvan, D. et al., 2001, A\&A 377, 343
              
\bibitem[1993]{cruikshank93}Cruikshank, D.P. et al. 1993, Science, 261, 742

%\bibitem[1992]{delitsky92}Delitksy, M. 1992
\bibitem[1999]{doute}Dout\'e, S. et al. 1999, Icarus, 142, 421
 
\bibitem[1998]{elliot98}Elliot, J. L. et al. 1998, Nature 393, 765

\bibitem[2000]{elliot00}Elliot, J. L. et al. 2000a, Icarus, 148, 347
\bibitem[2000]{elliot00b}Elliot, J. L. et al. 2000b, Icarus, 143, 425

% \bibitem[2003]{elliot03}Elliot, J.L. et al. 2003, Nature 424, 165

%\bibitem[2007]{elliot07} Elliot, J.L. et al. 2007, Astron J.  134, 1-13

%\bibitem[2005]{fiorenza} Fiorenza, C. \& Formisano, V. 2005, Planet. Space Sci., 53, 1009

%\bibitem[1999]{Forget99}Forget, F., N. Decamp, and F. Hourdin 1999. A 3-D General Circulation Model
%  of Triton’s Atmosphere and Surface. Pluto and Triton: Comparisons and
%  Evolution over Time. Lowell Observatory, Flagstaff, AZ.

\bibitem[2003]{forget}Forget, F, et al., 2000, Bull. Amer. Astron. Soc., 32, 45.01.

\bibitem[2010]{Fray}Fray, N. \& Schmitt, B., 2010, Planet. Space. Sci., in press.

\bibitem[2010]{Grundy 2010}Grundy W. M. et al. 2010, Icarus, in press.

\bibitem[1998]{gurrola95}Gurrola, E.M., 1995, Ph.D. thesis, Stanford University, Palo Alto.

\bibitem[1992]{hansen92}Hansen, C.A. \& Paige, D.A., 1992, Icarus 99, 273
\bibitem[1991]{herbert91}Herbert, F;, \& Sandel, B.R., 1991, JGR, 96, 19241

% \bibitem[1988]{hubbard} Hubbard, W.B., et al. 1988, Nature 336, 452

 \bibitem[2004]{kaufl}K\"aufl, H.U. et al. 2004, SPIE, 5492, 1218

\bibitem[1993]{krasno93}Krasnopolsky, V.A., et al. R.J., 1993, JGR 98, 3065
\bibitem[1995]{krasno95}Krasnopolsky, V.A., \& Cruikshank, D.P., 1995, JGR 100, 21271

%\bibitem[1992]{lellouch92}Lellouch, E. 1992, Adv. Space Res.

%\bibitem[2000]{lellouch00}Lellouch, E. et al. 2000, Icarus, 147, 220

\bibitem[2009]{lellouch09}Lellouch, E. et al., 2009, A \& A,  495, L17

%\bibitem[1990]{margolis} Margolis, J.S. 1990 Appl. Opt. 29, 2295

\bibitem[2007]{olkin}Olkin, C.B. et al. 1997, Icarus,  129, 178

\bibitem[1983]{prokh}Prokhvatilov, A. I.\& Yantsevich, L.D. 1983, Sov. J. Low Temp. Phys. 9,
  94, 

\bibitem[1999]{quirico}Quirico, E., et al. 1999, Icarus 139, 159

\bibitem[2003]{sicardy98}Sicardy, B. et al. 1998, Bull. Amer. Astron. Soc., 30, 49.02.

\bibitem[1992]{spencer92}Spencer, J.R. \& Moore, J. M. 1992, Icarus 99, 261

\bibitem[1997]{spencer}Spencer, J.R., et al. 1997, in Pluto and Charon,  eds. S.A. Stern \& D.J. Tholen  (The University of Arizona Press), 435

%\bibitem[1989]{stansberry89} Stansberry, J.A., Lunine, J.I. \&Tomasko, M.G. 1989, GRL, 16, 1221

%\bibitem[1994]{stansberry94} Stansberry, J.A., et al. 1994, Icarus, 111, 503

\bibitem[1996]{stansberry96} Stansberry, J.A. et al. 1996, Planet. Space Sci. 44, 1051

%\bibitem[1992]{stevens}Stevens, 1992
\bibitem[1997]{strobel95}Strobel, D.F. \&  Summers, M.E. 1995, in Neptune and Triton, ed.  D.P. Cruikshank (The University of Arizona Press), 1107

\bibitem[1996]{strobel}Strobel, D.F., Zhu, X. \&  Summers, M.E. 1996, Icarus, 120, 266

%\bibitem[1997]{tholen}Tholen, D. J.\& Buie, M.W. 1997, 
%in Pluto and Charon, eds. S.A. Stern\& D.J. Tholen (The University of Arizona Press), 193

%\bibitem[1997]{trafton97}Trafton, L.M., Hunten, D.M., Zanhle, K.J. \& McNutt, R.L. 1997, in Pluto and Charon, eds.  S.A. Stern \& D.J. Tholen %(TheUniversity of Arizona Press), 475

\bibitem[1990]{trafton90}Trafton, L.M., 1990, Ap J, 359, 512

\bibitem[1998]{trafton98}Trafton, L.M, Matson, D.L.\& Stansberry J.A. 1998, in Solar System Ices, eds. B. Schmitt, C. de Bergh, and M. Festou
(Kluwer Academic Publishers), 773

\bibitem[1994]{tryka}Tryka, K. et al. 1994, Icarus, 112, 513

%\bibitem[1989]{yelle} Yelle, R.V.\& Lunine, J.I. 1989, Nature, 399, 288

\bibitem[1997]{yelle95}Yelle, R.V., Lunine, J.I., Pollack, J.B.,\& Brown, R.H., 1995, in Neptune and Triton, ed.  D.P. Cruikshank (The University of Arizona Press), 1031.

\bibitem[1997]{young97}Young, L.A., et al. 1997, Icarus, 127, 258

\bibitem[2001]{young01} Young, L.A. et al. 2001, Icarus, 153, 148

% \bibitem[2008]{lyoung08}  Young, L. et al. 2008, Bull. Amer. Astron. Soc. 40, 461


 \end{thebibliography}
\end{document}